\documentclass{aamas2016_arxiv}

\usepackage{cite}

\usepackage{enumitem}
\setlist[itemize]{leftmargin=*}

\usepackage{amsmath}
\usepackage{amsfonts}
\usepackage{verbatim}
\usepackage{hhline}
\usepackage{multirow}

\usepackage{pdfpages}
\usepackage{graphicx}
\usepackage{subcaption}

\usepackage{xspace}

\newcommand{\nats}{\mathbb N}

\newcommand{\eps}{\ensuremath{\epsilon}\xspace}

\newcommand{\profx}{\ensuremath{\mathbf{x}}\xspace}
\newcommand{\profy}{\ensuremath{\mathbf{y}}\xspace}

\newcommand{\PPAD}{\texttt{PPAD}\xspace}
\newcommand{\NP}{\texttt{NP}\xspace}
\newcommand{\PLS}{\texttt{PLS}\xspace}

\newcommand{\lemke}{\textsc{Lemke}\xspace}
\newcommand{\descent}{\textsc{Descent}\xspace}

\begin{document}

\title{An Empirical Study on Computing Equilibria\\ in Polymatrix Games}

\numberofauthors{4}

\author{
\alignauthor Argyrios Deligkas\\
    \affaddr{University of Liverpool, UK}
    %\affaddr{United Kingdom}
    %\email{a.deligkas@liverpool.ac.uk} 
\alignauthor John Fearnley\\
    \affaddr{University of Liverpool, UK}
    %\affaddr{United Kingdom}
    %\email{john.fearnley@liverpool.ac.uk} 
\and
\alignauthor Tobenna Peter Igwe\\
    \affaddr{University of Liverpool, UK}
    %\affaddr{United Kingdom}
    %\email{t.igwe@liverpool.ac.uk}
\alignauthor Rahul Savani\\
    \affaddr{University of Liverpool, UK}
    %\affaddr{United Kingdom}
    %\email{rahul.savani@liverpool.ac.uk}
}

\maketitle

\begin{abstract}
The Nash equilibrium is an important benchmark for behaviour in systems of strategic autonomous agents.  Polymatrix games are a succinct and expressive representation of multiplayer games that model pairwise interactions between players.  The empirical performance of algorithms to solve these games has received little attention, despite their wide-ranging applications.  In this paper we carry out a comprehensive empirical study of two prominent algorithms for computing a sample equilibrium in these games, Lemke's algorithm that computes an exact equilibrium, and a gradient descent method that computes an approximate equilibrium.  Our study covers games arising from a number of interesting applications.  We find that Lemke's algorithm can compute exact equilibria in relatively large games in a reasonable amount of time.  If we are willing to accept (high-quality) approximate equilibria, then we can deal with much larger games using the descent method.  We also report on which games are most challenging for each of the algorithms.
\end{abstract}

% Note that the category section should be completed after reference to the ACM Computing Classification Scheme available at
% http://www.acm.org/about/class/1998/.

%\category{J.4}{Social and Behavioral Sciences} {Economics}

%A category including the fourth, optional field follows...
%\category{D.2.8}{Software Engineering}{Metrics}[complexity measures, performance measures]

%General terms should be selected from the following 16 terms: Algorithms, Management, Measurement, Documentation, Performance, Design, Economics, Reliability, Experimentation, Security, Human Factors, Standardization, Languages, Theory, Legal Aspects, Verification.

\terms{Algorithms, Economics}

%Keywords are your own choice of terms you would like the paper to be indexed by.


\section{Introduction}

In multiagent systems it is often the case that autonomous agents interact with
each other, but do not necessarily have the same objectives or goals. This
situation can be described as a game played between the agents, and the tools
from game theory can be used to analyse the possible outcomes. In particular,
the concept of a Nash equilibrium~\cite{N} describes a stable situation in which
no agent can increase its reward by changing its behaviour. Therefore, to
gain insight into the possible behaviours that a system of rational agents will
produce, one can compute the Nash equilibria of the game that is played between
the agents.

Games are often represented in \emph{strategic-form}, where for each possible
combination of strategy choices, a numerical payoff is specified
for each of the players. However, the size of this representation grows
\emph{exponentially} in the number of players. For example, for a game with $n$
players who can each choose between $2$ strategies, $n \cdot 2^n$ payoffs must
be specified. Hence, the strategic-form is typically unsuitable for the types of
games that arise from multiagent systems.

Many realistic scenarios do not need the flexibility that strategic-form
games provide. In particular, it is often the case that only the \emph{pairwise}
interactions between players are important. In this paper we study
\emph{polymatrix games} which model this. In these games the
interaction between the players is specified as a graph. Each player plays
an independent two-player game against each other player that he is connected
to, and the same strategy must be played in all of his
games. A player's payoff is then the sum of the payoffs from each of the
games. Crucially, the representations of these games grow \emph{quadratically}
in the number of players, which makes them more suitable for representing the large
multiagent systems that arise from real-world scenarios.

While there has been a large amount of theoretical work on polymatrix
games~\cite{DGP,CDT,EY10,FT-C10,CPY13,GW04}, the practical aspects of computing
equilibria in polymatrix games have yet to be studied. In this paper, we
provide an empirical study of two prominent methods for computing equilibria in
these games. Firstly, we study \emph{Lemke's algorithm}. The Lemke-Howson 
algorithm is a
famous algorithm for finding Nash equilibria in bimatrix
games~\cite{lemke64}, and Lemke's algorithm is a more general
technique for solving
\emph{linear complementarity problems} (LCPs). Miller and Zucker \cite{miller91}
have shown that the problem of finding a Nash equilibrium in a polymatrix game
can be reduced to the problem of solving an LCP that can then be tackled by
Lemke's algorithm.

%Apart from brute force
%support enumeration (which is exponential for polymatrix games even for finding
%pure Nash equilibria), Lemke's algorithm is the only known method for finding an
%Nash equilibrium in a polymatrix game.

Secondly, we study a method for finding \emph{approximate} equilibria in
polymatrix games. In contrast to a Nash equilibrium, where no player has an
incentive to deviate, an approximate equilibrium allows the players to have a
positive, but small, incentive to deviate from their current strategies.
Approximate equilibria have received a large amount of interest in theoretical
work~\cite{BBM10,BGR08,DFSS,DMP,HRS08,TS}, because the problem of finding an
exact Nash equilibrium in a polymatrix game, even for only two players, is known
to be \PPAD-complete (which implies that there is unlikely to be a
polynomial-time algorithm for this problem). From a practical point of view, it
is reasonable to use sufficiently accurate approximate equilibria to study
real-world systems, because often there is a non-negative cost to changing
strategy, which could deter an agent from deviating even if doing so would lead
to a small increase in payoff. Also, if the game is derived from real data, then
any uncertainty in the actual payoffs of the underlying situation means that
agents may be perfectly happy in the real world, even if the game model says
that they can gain a small amount through deviation.

We study a recently proposed gradient descent-like algorithm for finding
approximate equilibria in polymatrix games~\cite{DFSS}, which is the only known
approximation technique for (general) polymatrix games. It generalizes
the algorithm of Tsaknakis and Spirakis (TS) for bimatrix games\cite{TS}.
A recent
study found that the TS algorithm typically finds high-quality approximate
equilibria in practice~\cite{FIS15}, much better that its
theoretical worst-case performance.

\medskip

\noindent \textbf{Our contribution.}
We provide a thorough empirical study of finding exact and
approximate Nash equilibria in polymatrix games. We develop an extensive library
of game classes that cover a number of applications of polymatrix games
including cooperation games, strictly competitive games, and
group-wise zero-sum games. We also study \emph{Bayesian two-player games}, 
which can be modelled as
polymatrix games. In particular, we focus on various forms of \emph{Bayesian
auctions} (e.g., item bidding combinatorial auctions) and Bayesian variants of
Colonel Blotto games, which have applications to task allocation and resource
allocation problems between agents~\cite{SL08}.
All of our algorithm implementations and
game generators are open source and publicly
available\footnote{\texttt{http://polymatrix-games.github.io/}}, so that any new
algorithms developed for polymatrix games can be tested against our test suite.

We study Lemke's algorithm and the descent method on all of the
problems that we consider. In total we applied Lemke's algorithm to 188,000
instances using 26 months of CPU time, while we applied descent to
213,000 instances using 2.7 months of CPU time.  We found that Lemke's algorithm
can compute exact equilibria in relatively large games in a reasonable amount of
time, though the descent method is much more scalable and can be used to compute
approximate equilibria for instances that are an order of magnitude bigger.
Moreover, in contrast to its
theoretical worst-case performance guarantee, the descent method typically finds
very high quality approximate equilibria.

%The contributions of this paper can be be divided into three main categories:
%implementation of the algorithms, construction and implementation of game 
%generators, experimental results. To our knowledge, this is the first 
%implementation of the \descent algorithm, and since it is not known whether the 
%analysis of the approximation guarantee is tight, our results give lower bounds 
%for the algorithm. 

%The second contribution of our project is the construction and the implementation
%of generators for two bidder Bayesian auctions, Bayesian blotto, and several 
%classes of polymatrix games with known and unknown complexity. Our plan is to 
%make the generators (and the algorithms' code) public to the community, 
%so they can
%can be used in the future as a test suite for other algorithms for polymatrix 
%and two player Bayesian games. Furthermore, the games generators we constructed 
%are rather general and easy to adapt in order to produce many different games. 
%To our knowledge these are the first generators only for polymatrix games. 
%GAMUT produces polymatrix games and then converts them into general n-player 
%games without giving the option to the end user to output a polymatrix game.  

%The last contribution is the experiments. Since the majority of the studies for
%the games we consider are theoretical, our experiments are an important first 
%step to understand better the average complexity for the known to be hard problems
%and to get some intuition for the games whose complexity remains unknown. 

%\vskip -\baselineskip
%\paragraph{\textbf{Related work}}
\medskip
\noindent \textbf{Related work.}
The problem of equilibrium computation has received much attention from the 
theoretical point of view. Firstly, it was proven that computing an exact Nash 
equilibrium is \PPAD-complete~\cite{DGP,CDT}, even for games with only 
two players. 
While the class \NP captures 
decision problems, the complexity class \PPAD captures problems where it is 
\emph{known} that a solution exists. It is assumed that it is unlikely that
there exists a polynomial time algorithm for \PPAD-complete problems.
For this reason a line of work that studies approximate 
notions of Nash equilibria has arisen~\cite{BBM10,BGR08,DMP,HRS08,TS}. 
Specifically for polymatrix games, there is the recent descent procedure 
studied in this paper~\cite{DFSS}, and a recent QPTAS for polymatrix games 
on trees~\cite{BLP15}.
%
%Polymatrix games have played a central role in the reductions that have 
%been used to show \PPAD-hardness of games and other equilibrium 
%problems~\cite{DGP,CDT,EY10,FT-C10,CPY13}. 

There are empirical studies on equilibrium computation both for exact 
equilibria~\cite{SGC05,ARSvS,ABH,PNS,FIS15} and approximate equilibria~\cite{FIS15}, 
but none of them focused on polymatrix games. Instead, these studies mainly 
focused on games created by GAMUT~\cite{GAMUT}, the most famous suite of game generators.
GAMUT has a generator for some simple polymatrix games, but it converts them to strategic-form
games which blows up the representation exponentially.

Polymatrix games have received a lot of attention recently.
Computing a Nash equilibrium in a polymatrix game is \PPAD-hard even when all
the bimatrix games are either zero-sum or coordination games~\cite{CD11}.
Recently, it was proven that there is a constant $\eps>0$ such that it is
\PPAD-hard to compute an \eps-Nash equilibrium of a polymatrix
game~\cite{Rub15}.
Govindan and Wilson proposed a 
(non-polynomial-time) algorithm for computing equilibria of an $n$-player 
strategic-form game, by approximating the game with a sequence of polymatrix games~\cite{GW04}.  
Later, they presented a (non-polynomial) reduction that reduces $n$-player games
to polymatrix games while preserving approximate Nash equilibria \cite{GW10}.
%Their reduction introduces a central coordinator player, who interacts
%bilaterally with every player.

Many papers have derived bounds on the Price of Anarchy~\cite{BR11,R12,FHGL} in
item bidding auctions~\cite{CKS08}.
Only recently Cai and Papadimitriou~\cite{CP14} and Dobzinski, Fu and
Kleinberg~\cite{DFK} studied the question of the complexity of the equilibrium
computation problem in this setting. Blotto games are a basic model of resource
allocation, and have therefore been studied in the agents
community~\cite{PTW14,AHLMS16}. There have also been several papers that
study Blotto games with incomplete information as well, see for
example~\cite{AM09,KR11}.

Polymatrix games are examples of \emph{graphical games}, which are succinct
representations of games where interactions between players are encoded in a graph.
A related succinct representation is that of Action Graph Games (AGGs); 
introduced by Bhat and Leyton-Brown, AGGs capture local dependencies as
in graphical games, and partial indifference to other agents' identities as in
anonymous games~\cite{JLB11,BL04,DSVV09}.

%%%%%%%%%%%%%%%%%%%%%%%%%%%%%%%
%%%%%%%%%%%%%%%%%%%%%%%%%%%%%%%
%%%%%%%%%%%%%%%%%%%%%%%%%%%%%%%
\section{Preliminaries}
\label{sec:pre}
%\paragraph{\textbf{Bimatrix Games}}
\noindent \textbf{Bimatrix games.}
A bimatrix game is a pair $(R,C)$ of two $n \times n$ matrices: $R$ gives
payoffs for the \emph{row} player, and $C$ gives the payoffs for the
\emph{column} player. Each player has $n$ \emph{pure} strategies. To play the
game, both players simultaneously select a pure strategy: the row player selects
a row $i$, and the column player selects a column $j$. The row player then
receives $R_{i,j}$, and the column player $C_{i,j}$.

A \emph{mixed strategy} is a probability distribution over $[n]$. We denote a
row player mixed strategy as a vector $\profx$ of length $n$, such that
$\profx_i$ is the probability assigned to row $i$. Mixed strategies of the
column player are defined symmetrically. If $\profx$ and $\profy$ are mixed
strategies for the row and the column player, respectively, then the expected
payoff for the row player under the strategy profile $(\profx, \profy)$ is given
by $\profx^T R \profy$ and for the column player by $\profx^T C \profy$.

%\vskip -\baselineskip
%\paragraph{\textbf{Polymatrix Games}}
\medskip
\noindent \textbf{Polymatrix games.}
An $n$-player polymatrix game is defined by an $n$-vertex graph. Each 
vertex represents a player. 
Each edge $e$ corresponds to a bimatrix game that will be
played by the players that $e$ connects. Hence, a player with degree $d$ plays
$d$ bimatrix games. More precisely, each player picks a strategy
$x_i$ and plays that strategy in \emph{all} of the bimatrix games that he is
involved in. His expected payoff is given by the sum of the expected payoffs that
he obtains over all the bimatrix games that he participates in. We use $\profx = (x_1,
\ldots, x_n)$ to denote a strategy profile of an $n$-player game, where $x_i$
denotes the mixed strategy of player $i \in [n]$.

%\vskip -\baselineskip
%\paragraph{\textbf{Solution concepts}}
\medskip
\noindent \textbf{Solution concepts.}
The standard solution concept for strategic-form games is the \emph{Nash equilibrium} 
(NE). A relaxed version of this concept is the approximate NE, or \eps-NE. 
Intuitively, a strategy profile is an \eps-NE in an $n$-player game, if no player 
can increase his utility more than \eps by unilaterally changing his strategy. 
To put it formally, let \profx denote a strategy profile for the players and let
$u_i(z, \profx_{-i})$ denote the utility of player $i$ when he plays the 
strategy $z$ and the rest of the players play according to $\profx$. We say that
\profx is an \eps-NE if for every player $i$ it holds that
%\begin{align*}
$u_i(x_i, \profx_{-i}) \geq u_i(z, \profx_{-i}) - \eps$ for all possible $z$.
%\end{align*}
If $\eps = 0$, we have an exact Nash equilibrium.
%The same notion of approximation can be applied for Bayesian Nash equilibria 
%(\eps-BNE). The reduction presented in~\cite{DFSS} is approximation preserving,
%that is every $\eps$-NE in the polymatrix game constructed form a Bayesian game 
%$B$, is an $\eps$-BNE in $B$ too.

%%%%%%%%%%%%%%%%%%%%%%%%%%%%%%%%%%%%%%%%%%%%%%%%%%%%%%%%%%%%%%%%%%%%%%%%%%%%%%%%
%%%%%%%%%%%%%%%%%%%%%%%%%%%%%%%%%%%%%%%%%%%%%%%%%%%%%%%%%%%%%%%%%%%%%%%%%%%%%%%%
%%%%%%%%%%%%%%%%%%%%%%%%%%%%%%%%%%%%%%%%%%%%%%%%%%%%%%%%%%%%%%%%%%%%%%%%%%%%%%%%
\section{Algorithms}
%In this section, we describe the two algorithms that we study in this paper. 

%The first algorithm 
%computes exact Nash equilibria in polymatrix games albeit in exponential time in 
%in the input size in the worst case, whereas \descent computes in polynomial
%time, in the input size, approximate equilibria with constant approximation 
%guarantee.
%with theoretical approximation guarantee at least 0.5-NE.

%\vskip -\baselineskip
%\paragraph{\textbf{Lemke}}
\medskip
\textbf{Lemke's algorithm} is a complementary pivoting algorithm 
for the Linear
Complementarity Problem (LCP)~\cite{lemke65}. 
%An LCP is given by an $n \times n$
%matrix $M$ and a vector $q$ of size $n$. The problem is to find $w, z$ such
%that:
%\begin{align*}
 %w = q + Mz, \qquad w \ge \mathbf{0}, \qquad z \ge \mathbf{0}, \qquad z^\mathtt{T} w = 0.
%\end{align*}
Miller and Zucker have shown that finding a Nash equilibrium in
a polymatrix game can be reduced in polynomial time to a Lemke-solvable 
LCP~\cite{miller91}. 
So, we first turn the polymatrix game into an LCP, and then apply Lemke's
algorithm.  We shall refer to this algorithm as \lemke.

One drawback is that the reduction assumes a complete
interaction graph even if the actual interaction graph is not complete (by
padding with all-zero payoff bimatrix games).  Hence, for sparse polymatrix games the
reduction introduces a significant blowup, which affects the
performance of the algorithm.  To make this blowup clear, in our results we
report the number of payoffs in the original polymatrix game and the number of
matrix entries in the resulting LCP.

\medskip
\textbf{Descent} is a gradient descent-like algorithm proposed in~\cite{DFSS}.
It tries to minimize the \emph{regret} that a player suffers, which
is the difference between the best-response utility
and the actual utility he gets. 
%For a given strategy profile, each player will have a different regret,
%and the algorithm aims to reduce the regret of the player with the maximum
%regret.
%
The algorithm starts from an arbitrary strategy profile~\profx and in each
iteration it computes a new profile in which the maximum regret (over the
players) has been reduced.
%
%$\profx'$ which specifies the direction in which the
%maximum regret decreases with the highest rate. Then it produces a new
%strategy profile which is obtained by moving in this direction.
%
The algorithm takes a parameter $\delta$ that controls how accurate the
resulting approximate Nash equilibrium is. The theoretical results state that
the algorithm finds a $(0.5+\delta)$-NE after $O(\delta^{-2})$ iterations (for a fixed 
size game).
We test the cases where $\delta$ is either $0.1$ or
$0.001$, and we provide full results for both cases. We shall refer to
this algorithm as \descent in our results.

For the strategy profile
$\profx$, the algorithm first computes a direction vector $(\profx' - \profx)$,
and then moves a certain distance in that direction. In other words, the
algorithm moves to a new strategy profile $\profx + \alpha(\profx' - \profx)$,
where $\alpha$ is some constant in the range $(0, 1]$. The theoretical analysis
in~\cite{DFSS} uses $\alpha = \frac{\delta}{\delta + 2}$ in the proof of 
polynomial-time convergence. In practice we found that using
larger step sizes greatly improves the running time.
Hence, we adopt a \emph{line search} technique, which we adapt from the
two-player setting~\cite{TSK08}. It checks a number of equally
spaced values for $\alpha$ in $[0,1]$, and selects the best
improvement that is found. We also 
include $\alpha = \frac{\delta}{\delta + 2}$ as an extra value in this check, so
that the theoretical worst-case running time is unchanged. In
our results, we check 201 different points for $\alpha$ in each iteration.  For
a justification of the reasonableness of this choice, see
Appendix~A in the full version of this paper~\cite{DFIS16}.

\section{Game classes}
\label{sec:classes}

%The main structural difference between these categories
%is the underlying graph. In the Bayesian two player games is a complete bipartite
%graph, whereas for the latter can be any arbitrary graph. 

%\subsection*{Bayesian two player games}

\subsection*{Bayesian Auctions}

%\vskip -\baselineskip
%\paragraph{\textbf{Combinatorial auctions}} 
\noindent \textbf{Combinatorial auctions.} 
In a combinatorial auction, $m$ items are auctioned to $n$ bidders. Each bidder 
has a valuation function that assigns a non-negative real number to every subset 
of the items. 
Notice that in this setting, in general, the size required to represent the 
valuation function is exponential in $m$.
In combinatorial auctions with \emph{item bidding}, each player bids for every
item separately and all the items are auctioned simultaneously. A bidder wins an
item if he submitted the highest bid for that item. The bidders pay
according to a predefined payment rule. We study three popular
payment rules.
In a \emph{first price} auction the winner of an item has to pay his bid for that item,
in a \emph{second price} auction the winner of an item has to pay the second highest
bid submitted for the item, 
and in an \emph{all pay} auction every bidder has to pay his bid 
\emph{irrespective} of whether he won the item or not.
If more than one bidder has the highest bid
for some item, we resolve this tie according to a predefined publicly known
rule. We study two tie-breaking rules: either we always favor one
of the players, or we choose the winner for each item independently uniformly at
random.

We say that a combinatorial auction allows \emph{overbidding} if a bidder is 
allowed to make a bid for a item greater than his value for it. 
A common assumption in the literature is that overbidding is not allowed, since
allowing overbidding leads to the existence of trivial equilibria. 
Therefore, in our experiments we do not allow overbidding.

In a \emph{Bayesian}  combinatorial auction the valuation
function for every player is chosen according to a commonly known joint probability
distribution, which in this paper is always discrete. 
The different valuation functions that may be drawn for a player are 
known as his~\emph{types}. 
%In this paper we study Bayesian combinatorial auctions with two
%bidders. 

\medskip

%\vskip -\baselineskip
%\paragraph{\textbf{Item bidding auctions}} 

\noindent \textbf{Item bidding auctions.} 
We create two-bidder Bayesian item bidding combinatorial auctions with 
2 to 4 items for sale and 2 to 5 different types per player.
Each player's type (valuation function) is chosen uniformly at random. 
We study several well known valuation functions:
\begin{itemize}
\itemsep0em
\item \emph{Additive}: the value of each bundle of items is the sum of the 
values of the items contained in the bundle.
\item \emph{Budget additive}: the value of each bundle is the minimum of a
	budget parameter and the sum of the values of the items contained in the
	bundle.
\item \emph{Single minded}: each bidder has positive value for a specific bundle 
of items (and the same value for any other bundle of items containing that
bundle) and zero otherwise.
\item \emph{Unit demand}: the value of each bundle is the maximum value the bidder
has for any single item contained in the bundle.
\item \emph{AND-OR}: the first bidder has positive value only for the grand bundle
of items and zero otherwise, while the second one has a unit demand valuation.
\end{itemize}
%
%Since we study Bayesian auctions with item bidding, a pure strategy for a bidders 
%when $n$ items are auctioned is defined as $(b_1, \ldots, b_n)$ where $b_i$ 
%represents the bid the player made for item $i$. We allow bid increments of one
%unit for every $b_i$. In all the created games we do 
%not allow overbidding; as mentioned earlier too, there are trivial equilibria if
%overbidding is allowed. A player overbids for a subset of items if the sum of 
%his bids for the items in this subset is greater than his valuation for the 
%subset. 
To create the valuations, we set a maximum value~$M\in \nats$ that a player 
can have for any item, and a minimum value $m \in \nats$ such that in every valuation
there must be an item with a value of at least $m$. Bids are restricted to $\nats$, so 
$M$ and $m$ define the number of pure strategies a player has, and
consequently the size of the game. For a player with a single-minded
valuation function, we choose a random subset of items and a random value for
that subset in the range $[m, M]$. For additive and unit demand valuations, we
randomly select a value for each item from the set of allowed valuations. The
same procedure is extended for budget additive bidders: first we draw values for
items as for additive bidders; then we draw the budget as a random integer
in $[M,N]$ where~$N$ is the sum of the values for the items.

%\vskip -\baselineskip
%\paragraph{\textbf{Multi-unit auctions}}

\medskip

\noindent \textbf{Multi-unit auctions.}
In these auctions all the items being sold are 
identical. When there are $n$ items for sale, a valuation is 
given by an $n$-tuple $(v_{1}, \ldots, v_{n})$, where $v_{j}$ represents the 
player's marginal value for receiving a $j$-th copy of the item. Hence, the 
valuation for a bidder when he wins~$k$ items is the sum of the values $v_1$ up 
to $v_k$. 
%
%A pure strategy for a player is an $n$-tuple of bids in non-increasing order. 

Again we study the three most common payment rules: the first price rule, a.k.a.\
the \emph{discriminatory auction}, where a player that won $k$ items has to pay the
sum of his $k$ highest bids; the second price rule, a.k.a.\ the \emph{uniform-price
auction}, where the price for every item is the market-clearing price, i.e., the
highest losing bid; and the all-pay rule, where a player has to pay the sum of
his bids. 

We consider two well known valuation functions:
additive, where $v_j = v_1$ for all $j > 1$, and submodular, where $v_{j} \ge
v_{j+1}$ for all $j \in [n-1]$. We create games with 2 to 4 items
and 2 to 5 different types per player. The sampling of non-additive 
sub-modular valuation functions is not described here; we refer the reader to
the source code for further details\footnote{\texttt{http://polymatrix-games.github.io/}}.

%%%%%%%%%%%%%%%%%%%%%%%%%%%%%%%%%%%%%%%%%%%%%%%%%%%%%%%%%%
\section*{Other Bayesian Two-player Games}
%\vskip -\baselineskip
%\paragraph{\textbf{Two-Player Bayesian Games}}
A two-player Bayesian game is played between a row player and a column player. 
Each player has a set of possible types, and at the start of the game, each 
player is assigned a type according to a publicly known joint probability 
distribution. Each player learns his type, but not the type of the other player.
Rosenthal and Howson showed that the problem of finding an exact equilibrium in
a two-player Bayesian game can be reduced to finding an exact equilibrium in a
polymatrix game~\cite{HR74}, and this was extended to 
approximate equilibria in~\cite{DFSS}.
The underlying graph in the resulting polymatrix game is a complete 
bipartite graph where the vertices of each side represent the types of a player.
More specifically, if the row player has $n$ types and the column player has $m$
types, the corresponding polymatrix game has $n+m$ vertices and the payoff matrix
for edge $(uv)$ corresponds to the payoff matrix of the Bayesian game where the
row player has type $u$ and the column player has type $v$. We study the
following Bayesian two-player games.

\medskip

%\vskip -\baselineskip
%\paragraph{\textbf{Colonel Blotto games}}

\noindent \textbf{Colonel Blotto games.}
In Colonel Blotto games, each player has a number of soldiers $m_1$, $m_2$ that are 
simultaneously assigned to $n$ hills. 
Each player has a value for each hill that he receives if he 
assigns strictly more soldiers to the hill than his opponent and any ties are 
resolved by choosing a winner uniformly at random. The value that a player has 
for a hill is generated independently uniformly at random and the 
payoff a player 
gets under a strategy profile is given by the sum of the value of the hills 
won by that player. We consider games with 3 and 4 hills and 3 to 15
soldiers per player. We study two different Bayesian parameters: the valuations 
of the players over the hills and the number of soldiers that each player has. 
In the game we looked at, only one of these two parameters was used (i.e., for the 
other there was complete information).
For both cases we study games with 2 to 4 types per player.
When the types correspond to different valuations for hills, for every type, 
the valuations for each hill were drawn from independent uniform distributions on $[0, 1]$.
When the types correspond to the number of soldiers, we drew independently from
$\{3,\ldots,15\}$ for each player and each type.

\medskip

%\vskip -\baselineskip
%\paragraph{\textbf{Adjusted Winner games}}

\noindent \textbf{Adjusted Winner games.}
The adjusted winner procedure fractionally allocates a set of $n$ divisible items
to two players~\cite{AW}.  Under the procedure, $n-1$ items stay whole and
at most one is split between the players.  Each player has a
non-negative value for each item, and these values sum to 1.
Both players have additive valuations over bundles of items.  For a split item that 
a player has value $v$ for,
if the player receives $w \in [0, 1]$ of the item, he gets gets $w \cdot v$ value
from this part of the item.

The players simultaneously assign $m$ points to the items. Suppose player 1
assigns $\alpha_i$ points for the items $i=1,\ldots,n$ with $\sum_i \alpha_i =
m$, and similarly player 2's assignment is $(\beta_1, \dots,
\beta_n)$. The procedure starts with an initial allocation in which each item
goes to one of the players that assigned most points to it.  If the players get
equal utilities it stops. Otherwise it next determines which player gets higher
utility under this allocation, say player 1.  Next it finds the item~$i$ that is
currently allocated to player 1 and has the smallest ratio $\alpha_i / \beta_i$.
If possible it splits this item in a such a way as to equalize the total utilities of
the two players, or, if not, completely reallocates this item to player 2, and
repeats this step until the utilities of the two players are equalized.
Thus, at most one item is actually split.

We create Bayesian games where the players' types are different
valuations for the items.
We study the cases of 2 to 4 items and between 3 and 15 points for the players
to assign.  Independently for every type, the valuations for each item were drawn from independent
uniform distributions on $[0, 1]$, and then normalized to sum up to 1. 

\subsection*{Multi-player Polymatrix Games}
We study several types of game. For each, we study 
a range of underlying graphs: complete graphs, cycles, stars, and grid graphs.
In each case the entries of the payoff
matrices are drawn from independent uniform distributions on~$[0,1]$.
%
%\begin{description}
%\itemsep0em
%\item[Net coordination games.]
%\vskip -\baselineskip
%\paragraph{\textbf{Net coordination games}}

\medskip
\noindent \textbf{Net coordination games.}
In these games, every edge~$e$ corresponds to a coordination bimatrix game 
$(A_e, A_e)$. These games posses a pure NE, 
which is \PLS-complete to compute. The complexity of finding a (possibly non-pure)
exact equilibrium is in 
$\PLS \cap \PPAD$~\cite{CD11}.

%\item[Coordination/zero-sum games.]
%\vskip -\baselineskip
%\paragraph{\textbf{Coordination/zero-sum games}}

\medskip
\noindent \textbf{Coordination/zero-sum games.}
Here each edge is either a coordination or zero-sum game, i.e., on edge $e$ the
bimatrix game is $(A_e, A_e)$, or $(A_e, -A_e)$. These games are
\PPAD-complete~\cite{CD11} to solve.
We create games having a proportion $p$ of coordination games for $p \in \{0,
0.25, 0.5, 0.75, 1\} $. We study how $p$ affects the running time of
the algorithms. 

%\item[Group-wise zero-sum games.]
%\vskip -\baselineskip
%\paragraph{\textbf{Group-wise zero-sum games}}

\medskip
\noindent \textbf{Group-wise zero-sum games.}
The players are partitioned into groups so that the edges going between groups
are zero-sum while those within the same group are coordination games. In other
words, players inside a group are ``friends'' who want to coordinate their
actions, while players in different groups are competitors. These games are
\PPAD-complete~\cite{CD11} to solve even with 3 groups. 
We create games with 2 and 5 groups, all played on complete
graphs. In every case, each group is approximately the same size, 
and each player is assigned to a group at random. 

%\todo[inline]{
%The partition that a player belongs to is 
%chosen in the following way: we order the players and then we put each one of them to the
%group with number equal to his order modulo number of partitions. }

%\item[Strictly competitive games.]
%\vskip -\baselineskip
%\paragraph{\textbf{Strictly competitive games}}

\medskip
\noindent \textbf{Strictly competitive games.}
A bimatrix game is strictly competitive if for every pair of mixed strategy profiles $s$
and~$s'$ we have that: if the payoff of one player is better in $s$ than in $s'$, then the 
payoff of the other player is worse in $s$ than in $s'$. 
%These games were known 
%to be solvable via linear programming~\cite{aumann87}, and recent work has shown
%that they are merely affine transformations of zero-sum games~\cite{AdlerDP09}. 
%That is, if $(R,C)$ is a strictly competitive game, there exists a zero-sum game 
%$(R',C')$ and constants $c_1, c_2 > 0$ and $d_1, d_2$ such that $R = c_1R'+d_1I$
%and $C = c_2C'+d_2I$, where $I$ is the all-ones matrix. 
We study polymatrix games with strictly competitive games on the edges, which are
\PPAD-complete~\cite{CD11}.

\medskip

%\item[Weighted cooperation games.]
%\vskip -\baselineskip
%\paragraph{\textbf{Weighted cooperation games}}

\noindent \textbf{Weighted cooperation games.}
The unweighted version of these games was introduced in~\cite{ARSS15}. 
Each player chooses a colour from a set of available colours. The payoff of a
player is the number of neighbours who choose the same colour. These games have
a pure NE that can be computed in polynomial time. We study the more general
case where each edge has a positive weight. The complexity for the weighted case
is unknown~\cite{DP11}. We create games where all players have the same number
of available colors $k$, where $k$ is in $\{15,\ldots,45\}$.
For every player, his available colors are chosen uniformly at random
from all $k$-sized subsets from a universe of colors of size either~$2k$ or~$5k$.

%\todo[inline]{Will we include transduction games??}
%\item \textbf{Transduction games.}
%Graph transduction is a popular class of semi-supervised learning techniques which
%aims to estimate a classification function defined over a graph of labeled and 
%unlabeled data points. In~\cite{ErdemP12} the transduction problem is formulated 
%in terms of a polymatrix game where NE correspond to consistent labelings of the 
%data. The payoff matrix between players $i$ and $j$ is of the form 
%$A_{ij}=w_{ij}*I$ where $I$ is the identity matrix.

%\item \textbf{Ranking}
%
%\item \textbf{Security Games}
%In the security game which we studied, the defender is a security agent who wants to secure $k$
%houses, but can only patrol $d$ houses while the theif seeks to break into a house and steal it's
%valuables. If a theif chooses to rob a house which the security guard is not patrolling, then he is
%successful, otherwise, if he robs the $j$-th house in the patrol order, he is caught with
%probability $p_j$, such that for all $j$, $p_{j} \ge p_{j+1}$. If the theif successfully breaks in
%to a house, then he gains his valuation and the security guard loses theirs. In the event that the
%theif gets caught, he then pays a fine for getting caught and the guard gains a value for catching
%the theif.

%\end{description}

%%%%%%%%%%%%%%%%%%%%%%%%%%%%%%%%%%%%%%%%%%%%%%%%%%%%%%%%%%%%%%%%%%%%%%%%%%%%%%%%%
%%%%%%%%%%%%%%%%%%%%%%%%%%%%%%%%%%%%%%%%%%%%%%%%%%%%%%%%%%%%%%%%%%%%%%%%%%%%%%%%%
%%%%%%%%%%%%%%%%%%%%%%%%%%%%%%%%%%%%%%%%%%%%%%%%%%%%%%%%%%%%%%%%%%%%%%%%%%%%%%%%%
\section{Experimental setup}

%\vskip -\baselineskip
%\paragraph{\textbf{Implementation}}
%\noindent \textbf{Implementation.}
The algorithms and game generators were implemented in~C. The
CPLEX library was used for solving LPs in the implementation of \descent. 
Our implementation of \lemke uses integer pivoting
in exact arithmetic using the GMP library; we were unable to produce a
numerically stable floating point implementation
(generally our attempts would start to fail on LCP instances of dimension $60$).
%Due to floating point inaccuracies, we were unable
%to produce a stable implementation of the \lemke algorithm which worked on LCP's
%of size greater than $60 \times 60$ (an issue which plagues several libraries
%with an implementation of \lemke's algorithm). This usually lead to incorrect
%pivots being made within the impementation causing it to eventually ray
%terminate.
%
%\todo[inline]{Argy's comment: I believe that this explanation is not convincing 
%enough. What exactly are the inaccuracies and HOW they affect the algo? In the 
%way it is written someone can argue that we favoured \descent.}
%As a result, for \lemke's algorithm, we
%used an implementation which performs integer pivoting in exact arithmetic using the GMP library.
%
%For our runtime results, we only measure the amount of time which is spent
%running the algorithm.
%\todo[inline]{What does the above mean - we don't count the reduction to an
%LCP for Lemke? Anything else?}
All experiments had a time-out of 10 minutes. In our results, the average
runtime of the algorithms, as well as the approximation guarantee found,
include the instances which timed out. 
The experiments used a cluster of
8 machines with Intel Core i7-2600 CPU's clocked at 3.40GHz and 16GB of memory, running
Scientific Linux 6.6 with kernel version 2.6.32.

%%%%%%%%%%%%%%%%%%%%%%%%%%%%%%%%%%%%%%%%%%%
\section{Results}

\begin{table*}[htp]{
\centering

\begin{tabular}{|l|c|c|c||c|c|c||c|c||c|c|}
\multicolumn{4}{|c||}{Games} & \multicolumn{3}{c||}{\lemke} & \multicolumn{2}{c||}{\descent 0.1 LS} &
\multicolumn{2}{c|}{\descent 0.001 LS} \\\hline
& Valuation & Avg. Size & Auc & Time & \% Timeout & \% Pure & Time & \eps & Time & \eps \\
\hhline{|=|=|=|=||=|=|=||=|=||=|=|}
\multirow{15}{*}{\rotatebox[origin=c]{90}{Itembidding}}& \multirow{3}{*}{Additive} & \multirow{3}{*}{623990} & FP & 35.005 & 0.0 & 0.0 & 0.200 & 1.133e-02 & 1.287 & 2.630e-04\\\cline{4-11}
& & & SP & 0.764 & 0.0 & 100.0 & 0.233 & 7.103e-03 & 0.232 & 2.312e-05\\\cline{4-11}
& & & AP & 214.049 & 12.0 & 0.0 & 0.289 & 6.163e-03 & 1.610 & 1.943e-04\\
\hhline{|~|=|=|=||=|=|=||=|=||=|=|}
& & \multirow{3}{*}{650417} & FP & 46.351 & 0.0 & 34.0 & 0.199 & 6.249e-02 & 3.336 & 3.063e-02\\\cline{4-11}
& Unit &  & SP & 203.107 & 19.0 & 58.0 & 0.459 & 1.442e-02 & 1.831 & 2.647e-04\\\cline{4-11}
& &  & AP & 14.709 & 0.0 & 21.0 & 0.168 & 4.182e-02 & 2.757 & 1.113e-02\\
\hhline{|~|=|=|=||=|=|=||=|=||=|=|}
& & \multirow{3}{*}{519055} & FP & 280.322 & 23.0 & 0.0 & 0.194 & 1.927e-02 & 1.732 & 1.594e-04\\\cline{4-11}
& AndOr &  & SP & 1.269 & 0.0 & 100.0 & 0.279 & 7.026e-03 & 0.595 & 1.019e-04\\\cline{4-11}
& &  & AP & 166.011 & 9.1 & 0.0 & 0.171 & 8.662e-03 & 1.328 & 2.842e-04\\
\hhline{|~|=|=|=||=|=|=||=|=||=|=|}
& & \multirow{3}{*}{647583} & FP & 100.444 & 5.0 & 3.16 & 0.206 & 3.055e-02 & 2.003 & 4.064e-03\\\cline{4-11}
& Budget &  & SP & 9.438 & 1.0 & 84.85 & 0.348 & 2.543e-02 & 0.915 & 2.186e-04\\\cline{4-11}
& &  & AP & 248.739 & 24.0 & 0.0 & 0.271 & 1.990e-02 & 2.199 & 2.233e-03\\
\hhline{|~|=|=|=||=|=|=||=|=||=|=|}
& & \multirow{3}{*}{606511} & FP & 82.166 & 7.37 & 5.0 & 0.139 & 2.775e-02 & 1.551 & 1.822e-03\\\cline{4-11}
& SingleMinded &  & SP & 110.341 & 17.0 & 48.19 & 0.475 & 1.045e-02 & 1.359 & 1.669e-04\\\cline{4-11}
& &  & AP & 59.942 & 3.0 & 1.0 & 0.162 & 1.856e-02 & 1.544 & 1.038e-03\\
\hhline{|=|=|=|=||=|=|=||=|=||=|=|}
\multirow{6}{*}{\rotatebox[origin=c]{90}{Multiunit}}& \multirow{3}{*}{Additive} & \multirow{3}{*}{836465} & D 
& 9.504 & 0.0 & 30.0 & 0.295 & 1.452e-02 & 1.481 & 4.411e-04\\\cline{4-11}
& & & U & 0.954 & 0.0 & 100.0 & 0.380 & 1.321e-02 & 1.870 & 3.579e-04\\\cline{4-11}
& & & AP & 512.564 & 64.0 & 0.0 &  0.417 & 4.080e-03 & 2.182 & 3.170e-04\\
\hhline{|~|=|=|=||=|=|=||=|=||=|=|}
& \multirow{3}{*}{SubModular} & \multirow{3}{*}{878491}& D 
& 29.054 & 0.0 & 5.0 & 0.270 & 1.326e-02 & 1.954 & 3.209e-04\\\cline{4-11}
& & & U & 5.818 & 0.0 & 83.0 & 0.272 & 2.636e-02 & 1.741 & 8.645e-04\\\cline{4-11}
& & & AP & 210.290 & 12.0 & 0.0 & 0.323 & 1.115e-02 & 2.192 & 3.273e-04\\\hline\end{tabular}

\caption{Results for item bidding and multi-unit auctions with 3 items and 3
types per player. Ties are broken by
favouring the second player for item bidding, and by allocating the item uniformly at
random in the multi-unit case. For \lemke, we report the average running time, the percentage of instances
that exceeded our timeout of 10 minutes, and the percentage of instances for
which the algorithm finds a pure equilibrium. For \descent, we report the
average running time and the approximation quality of the approximate equilibrium that was found.}
\label{tbl:auction}
}
\end{table*}
\begin{table*}[htp]{
\centering
{\small

\begin{tabular}{|l|c|c||c|c||c|c||c|c|c|}
\multicolumn{3}{|c||}{Games} & \multicolumn{2}{c||}{\lemke} & \multicolumn{2}{c||}{\descent 0.1 LS} &
\multicolumn{3}{c|}{\descent 0.001 LS} \\\hline
Game & \# Types & \# Points/Troops& Time & \% Timeout & Time & \eps & Time & \eps & \% Timeout\\
\hhline{|=|=|=||=|=||=|=||=|=|=|}
AdjWinner & 30 & 5 & 281.407 & 21.0 & 0.939 & 4.450e-02 & 3.065 & 7.469e-03 & 0.0 \\\hline
AdjWinner & 3 & 60 & 233.963 & 10.0 & 138.584 & 2.683e-02 & 220.254 & 1.827e-03 & 10.0 \\
\hhline{|=|=|=||=|=||=|=||=|=|=|}
Blotto & 3 & 8 & 71.814 & 0.0 & 0.032 & 9.181e-03 & 0.480 & 5.281e-04 & 0.0 \\\hline
Blotto & 3 & 10 & 382.497 & 23.0 & 0.063 & 8.859e-03 & 0.845 & 5.408e-04 & 0.0 \\\hline
Blotto & 3 & 12 & 573.663 & 91.0 & 0.118 & 8.590e-03 & 1.392 & 6.268e-04 & 0.0 \\\hline
\end{tabular}

}
\caption{Results for Adjusted Winner and Blotto games. 
%The table shows the
%number of types and points for adjusted winner games, and the number of hills
%and troops for Blotto games. 
The two rows for Adjusted Winner show similar running
times but actually correspond to very different input sizes, with the second row 
corresponding to much larger games. The underlying reason is that the number of players 
in the polymatrix game (i.e., number of types) affects the running time
much more that the number of actions (i.e., number of items/troops). Also see
Appendix~B in the full version of this paper~\cite{DFIS16}.
}
\label{tbl:bayesian} 
} 
\end{table*}

% The next two tables are small and each fit in one column
%\begin{table*}[htp]
%\begin{subtable}{0.5\textwidth}{
%\centering
%\input{tables/auc_small/tables2.tex}
%\subcaption{Results for \lemke on first price (FP), second price (SP) and all-pay
%(AP) auctions with budget additive valuations, 3 items and 5 types per player.
%In the first two columns, all tied items are given to player 1,
%while in the last two, tied items are given uniformly at random
%}
%\label{tbl:ties}
%}
%\end{subtable}%
%\begin{subtable}{0.5\textwidth}{
%\centering
%\input{tables/auc_small/tables.tex}
%}
%\subcaption{Results for first price (FP), second price (SP),
%and all-pay (AP) auctions for additive bidders, with \lemke timing out on 13\% and 44.5\% of
%FP and AP auctions respectively, while \descent without line search times out on
%61.5\%, 56 \% and 94\% of instances on the respective auctions.}
%\label{tbl:ls}
%\end{subtable}
%\end{table*}

\begin{table*}[htp]{
\centering
\begin{tabular}{|c||c|c||c|c|}
\hline
Auc & \multicolumn{2}{c||}{Player 1} & \multicolumn{2}{c|}{Random}\\\hline
 & Time & \% Timeout & Time & \% Timeout \\
\hhline{|=||=|=||=|=|}
FP & 63.411 & 0.0 & 408.223 & 43.0 \\\hline
SP & 3.663 & 0.0 & 39.727 & 3.0 \\\hline
AP & 100.949 & 4.0 & 248.665 & 19.0 \\\hline
\end{tabular}

\caption{Results for \lemke showing the impact of the tie-breaking rule. We
report on first price (FP), second price (SP) and all-pay
(AP) auctions with budget additive valuations, 3 items, and 5 types per
player. In the first two columns, all tied items are allocated to player 1,
while in the last two, tied items are allocated uniformly at random.
}
\label{tbl:ties}
}
\end{table*}

\begin{table*}[htp]{ 
\centering
\begin{tabular}{|c||c||c|c||c|c|}
\hline
& \lemke & \multicolumn{2}{c|}{\descent 0.001} & \multicolumn{2}{|c|}{\descent LS 0.001}\\\hline
& Time & Time & \eps & Time & \eps \\\hline
FP & 229.4 & 508.0 & 5.4e-03 & 4.894 & 6.768e-04\\\hline
SP & 1.6 & 470.8 & 3.7e-03 & 0.491 & 1.013e-05\\\hline
AP & 547.3 & 496.0 & 6.5e-03 & 5.551 & 3.283e-04\\\hline
\end{tabular}

\caption{Results showing the impact of line search for \descent. We report
results for first price (FP), second price (SP),
and all-pay (AP) auctions for additive bidders. \lemke timed out on 13\% and 44.5\% of
FP and AP auctions respectively, while \descent without line search times out on
61.5\%, 56 \% and 94\% of instances on the respective auctions.}
\label{tbl:ls}
}
\end{table*}

\begin{table*}[t]{
\centering

\begin{tabular}{|c|c|c|c|c||c|c||c|c||c|c|c|}
\multicolumn{5}{|c||}{Games} & \multicolumn{2}{c||}{\lemke} & \multicolumn{2}{c||}{\descent 0.1 LS} &
\multicolumn{3}{c|}{\descent 0.001 LS} \\\hline
Game & Graph & \# Payoff & LCP & $p$ & Time & \% T & Time & \eps & Time & \eps & \% T\\
\hhline{|~|=|=|=|=||=|=||=|=||=|=|=|}
\multirow{20}{*}{\rotatebox[origin=c]{90}{Coord-Zero}}& \multirow{5}{*}{Complete} &
\multirow{5}{*}{26010} & \multirow{5}{*}{32400} & 0 & 1.270 & 0.0 & 0.034 & 2.103e-02 & 0.760 & 9.951e-04 & 0.0 \\\cline{5-12}
 & & & & 0.25 & 63.407 & 4.0 & 0.033 & 2.115e-02 & 0.748 & 1.026e-03 & 0.0 \\\cline{5-12}
 & & & & 0.5 & 337.443 & 45.0 & 0.034 & 1.859e-02 & 0.750 & 1.070e-03 & 0.0 \\\cline{5-12}
 & & & & 0.75 & 522.207 & 74.0 & 0.033 & 1.604e-02 & 0.725 & 1.076e-03 & 0.0 \\\cline{5-12}
 & & & & 1 & 116.354 & 0.0 & 0.034 & 4.844e-03 & 0.598 & 5.087e-04 & 0.0 \\
\hhline{|~|=|=|=|=||=|=||=|=||=|=|=|}
 & \multirow{5}{*}{Cycle} & \multirow{5}{*}{25920} & \multirow{5}{*}{136900} & 0 & 18.430 & 2.0 & 0.103 & 3.352e-02 & 3.612 & 1.093e-03 & 0.0 \\\cline{5-12}
 & & & & 0.25 & 184.451 & 21.0 & 0.103 & 3.157e-02 & 3.534 & 1.167e-03 & 0.0 \\\cline{5-12}
 & & & & 0.5 & 412.947 & 55.0 & 0.105 & 2.859e-02 & 3.430 & 1.136e-03 & 0.0 \\\cline{5-12}
 & & & & 0.75 & 593.414 & 96.0 & 0.103 & 2.626e-02 & 3.206 & 1.121e-03 & 0.0 \\\cline{5-12}
 & & & & 1 & 600.097 & 100.0 & 0.107 & 1.906e-02 & 2.712 & 6.557e-04 & 0.0 \\
\hhline{|~|=|=|=|=||=|=||=|=||=|=|=|}
 & \multirow{5}{*}{Grid} & \multirow{5}{*}{26136} & \multirow{5}{*}{93636} & 0 & 35.447 & 3.0 & 0.072 & 3.143e-02 & 2.257 & 1.023e-03 & 0.0 \\\cline{5-12}
 & & & & 0.25 & 260.455 & 35.0 & 0.069 & 3.239e-02 & 2.233 & 1.137e-03 & 0.0 \\\cline{5-12}
 & & & & 0.5 & 451.699 & 61.0 & 0.072 & 2.955e-02 & 2.254 & 1.170e-03 & 0.0 \\\cline{5-12}
 & & & & 0.75 & 552.286 & 82.0 & 0.072 & 2.786e-02 & 2.106 & 1.186e-03 & 0.0 \\\cline{5-12}
 & & & & 1 & 599.349 & 99.0 & 0.070 & 2.159e-02 & 1.802 & 6.489e-04 & 0.0 \\
\hhline{|~|=|=|=|=||=|=||=|=||=|=|=|}
 & \multirow{5}{*}{Tree} & \multirow{5}{*}{25992} & \multirow{5}{*}{152100} & 0 & 0.276 & 0.0 & 0.060 & 1.012e-02 & 0.818 & 1.175e-03 & 0.0 \\\cline{5-12}
 & & & & 0.25 & 0.542 & 0.0 & 0.062 & 1.997e-02 & 0.806 & 1.220e-03 & 0.0 \\\cline{5-12}
 & & & & 0.5 & 73.443 & 5.0 & 0.062 & 2.139e-02 & 0.814 & 1.246e-03 & 0.0 \\\cline{5-12}
 & & & & 0.75 & 165.418 & 4.0 & 0.061 & 2.150e-02 & 0.796 & 1.084e-03 & 0.0 \\\cline{5-12}
 & & & & 1 & 162.420 & 0.0 & 0.063 & 1.469e-03 & 0.778 & 7.686e-04 & 0.0 \\
\hhline{|=|=|=|=|=||=|=||=|=||=|=|=|}

 \multirow{9}{*}{\rotatebox[origin=c]{90}{Group Zero}}& \multirow{9}{*}{Complete} &
\multirow{3}{*}{20250} & \multirow{3}{*}{25600} & 2 & 368.032 & 36.0 & 0.025 & 1.976e-02 & 0.564 & 1.093e-03 & 0.0 \\\cline{5-12}
 & & & & 3 & 495.919 & 66.0 & 0.025 & 1.762e-02 & 0.550 & 1.129e-03 & 0.0 \\\cline{5-12}
 & & & & 5 & 435.207 & 29.0 & 0.025 & 1.308e-02 & 0.525 & 9.926e-04 & 0.0 \\\cline{3-12}
 & & \multirow{3}{*}{26010} & \multirow{3}{*}{32400} & 2 & 438.650 & 59.0 & 0.034 & 1.855e-02 & 0.760 & 1.068e-03 & 0.0 \\\cline{5-12}
 & & & & 3 & 576.439 & 91.0 & 0.034 & 1.583e-02 & 0.731 & 1.120e-03 & 0.0 \\\cline{5-12}
 & & & & 5 & 582.924 & 88.0 & 0.033 & 1.186e-02 & 0.677 & 9.738e-04 & 0.0 \\\cline{3-12}
 & & \multirow{3}{*}{36000} & \multirow{3}{*}{44100} & 2 & 545.997 & 84.0 & 0.052 & 1.564e-02 & 1.073 & 1.049e-03 & 0.0 \\\cline{5-12}
 & & & & 3 & 598.616 & 99.0 & 0.051 & 1.396e-02 & 1.037 & 1.110e-03 & 0.0 \\\cline{5-12}
 & & & & 5 & 600.088 & 100.0 & 0.051 & 1.101e-02 & 0.969 & 9.721e-04 & 0.0 \\
\hhline{|=|=|=|=|=||=|=||=|=||=|=|=|}

\multirow{4}{*}{\rotatebox[origin=c]{90}{Strict}}& Complete & 20250 & 25600& 5 & 356.009 & 17.0 & 0.024 & 1.878e-02 & 0.552 & 1.054e-03 & 0.0 \\\cline{2-12}
 & Cycle & 20480 & 108900 & 5 & 580.891 & 85.0 & 0.087 & 1.729e-02 & 2.102 & 1.068e-03 & 0.0 \\\cline{2-12}
 & Grid & 20184 & 72900 & 5 & 551.795 & 77.0 & 0.066 & 1.612e-02 & 1.428 & 1.108e-03 & 0.0 \\\cline{2-12}
 & Tree & 20808 & 122500 & 5 & 79.560 & 0.0 & 0.048 & 2.571e-03 & 0.664 & 8.111e-04 & 0.0 \\
\hhline{|=|=|=|=|=||=|=||=|=||=|=|=|}

 \multirow{12}{*}{\rotatebox[origin=c]{90}{Weighted Cooperation}}& \multirow{3}{*}{Complete} &
\multirow{3}{*}{995000} & \multirow{3}{*}{1440000} & 2 & 194.233 & 8.0 & 8.455 & 1.446e-02 & 86.116 & 9.603e-04 & 0.0 \\\cline{5-12}
 & & & & 3 & 410.118 & 38.0 & 6.551 & 1.909e-02 & 73.485 & 1.047e-03 & 0.0 \\\cline{5-12}
 & & & & 5 & 552.583 & 81.0 & 4.957 & 2.585e-02 & 64.676 & 1.130e-03 & 0.0 \\
\hhline{|~|=|=|=|=||=|=||=|=||=|=|=|}
 & \multirow{3}{*}{Cycle} & \multirow{3}{*}{17500} & \multirow{3}{*}{4410000} & 2 & 103.403 & 0.0 & 0.227 & 1.334e-01 & 577.984 & 3.094e-02 & 73.0 \\\cline{5-12}
 & & & & 3 & 90.062 & 0.0 & 0.172 & 1.412e-01 & 529.581 & 4.199e-02 & 48.0 \\\cline{5-12}
 & & & & 5 & 78.883 & 0.0 & 0.156 & 1.427e-01 & 438.707 & 3.950e-02 & 28.0 \\
\hhline{|~|=|=|=|=||=|=||=|=||=|=|=|}
 & \multirow{3}{*}{Grid} & \multirow{3}{*}{27200} & \multirow{3}{*}{3006756} & 2 & 116.157 & 0.0 & 0.864 & 8.298e-02 & 275.839 & 5.461e-03 & 1.0 \\\cline{5-12}
 & & & & 3 & 81.933 & 0.0 & 0.464 & 1.110e-01 & 480.608 & 1.368e-02 & 15.0 \\\cline{5-12}
 & & & & 5 & 58.054 & 0.0 & 0.131 & 1.384e-01 & 358.662 & 3.028e-02 & 2.0 \\
\hhline{|~|=|=|=|=||=|=||=|=||=|=|=|}
 & \multirow{3}{*}{Tree} & \multirow{3}{*}{24950} & \multirow{3}{*}{9000000} & 2 & 240.215 & 0.0 & 0.750 & 0.000e+00 & 2.768 & 3.253e-04 & 0.0 \\\cline{5-12}
 & & & & 3 & 220.510 & 0.0 & 0.709 & 0.000e+00 & 2.533 & 1.194e-04 & 0.0 \\\cline{5-12}
 & & & & 5 & 204.919 & 0.0 & 0.653 & 0.000e+00 & 2.345 & 8.947e-05 & 0.0 \\\hline\end{tabular}

\caption{Results for multi-player (non-Bayesian) polymatrix games.
The underlying graphs are complete graphs, cycles,
grids and star graphs. $\% T$ is the proportion of the timed out instances.
On Cooperation-Zerosum games, the value of $p$ represents the proportion of
games which are coordination games, for group zero-sum games, it represents the number of groups,
and for weighted cooperation games, it represents the multiplier dictating the total number of
colours available, i.e. if there are $k$ colours per player, then there are 
$k \cdot p$ total colours
available. }
\label{tbl:polymatrix}
}
\end{table*}

We ran both \lemke and \descent on all of our input instances.
Table~\ref{tbl:auction}
shows the results for auctions, Table~\ref{tbl:bayesian} shows the results for
other Bayesian two-player games, and Table~\ref{tbl:polymatrix} shows the results for
multi-player polymatrix games. While we tested games of many different sizes,
for the purposes of exposition, the tables 
display the largest instances that \lemke can solve without
timing out.

\begin{figure*}[h]
\centering
\includegraphics[page=1,width=\linewidth]{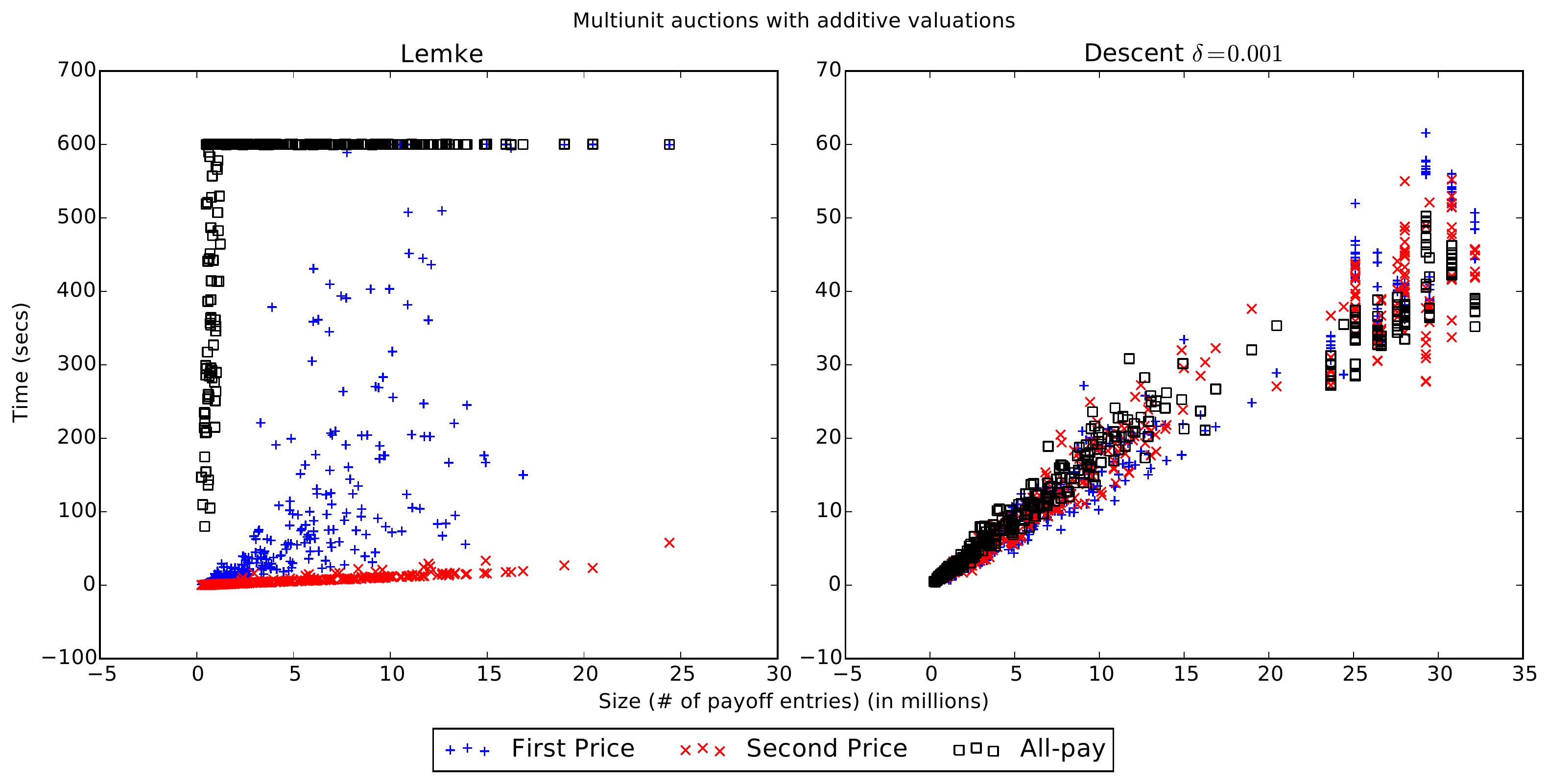}
\caption{
%Plots showing the performance \lemke and \descent with $\delta = 0.001$ on multi unit auctions on first price, second price
%and all-pay payment methods
Plots showing the performance of the algorithms on multi-unit auctions with
first price, second price, and all-pay payment rules. The left plot
shows the performance of \lemke's algorithm. It can clearly be seen that
the allocation rule impacts the performance of the algorithm. The right chart
right shows the performance of \descent with $\delta = 0.001$. The
$y$-axis scales on the two charts are not equal: \descent is much
faster than \lemke.
}
\label{fig:comparison}
\end{figure*}

One general feature of our results is that \descent is much
faster than \lemke. To illustrate this, Figure~\ref{fig:comparison} shows the
performance of the two algorithms on the three types of additive item-bidding
auctions included in the study. It can be seen that on the hard instances (first
price and all pay), \lemke starts to struggle when there are around 5 million
payoffs in the game, whereas even the slower and more accurate of the two
\descent variants ($\delta=0.001$) can handle games with 30 million payoffs in
under a minute.
Indeed, a runtime regression for Bayesian Blotto games
found that \lemke has roughly quadratic running time (with an
$R^2$ of $0.75$ for the regression), while \descent has roughly linear running time 
% 0.999 for delta 0.1 and 0.83 for delta 0.001
(with $R^2$s of $0.88$ and $0.96$ 
for the $\delta$ values of $0.1$ and $0.001$
respectively).

However, good runtime performance for \descent would be
of limited value if it only found poor quality approximate equilibria. Fortunately, our
results show that this is not the case. In almost all experiments \descent found
high quality approximate equilibria. The variant with $\delta = 0.1$
typically found an $\epsilon$-NE with $\epsilon \le 0.05$, while
the variant with $\delta = 0.001$ typically found an $\epsilon$-NE
with $\epsilon \le 0.002$.

\begin{figure*}[htp]
\centering
\includegraphics[page=1,width=\textwidth]{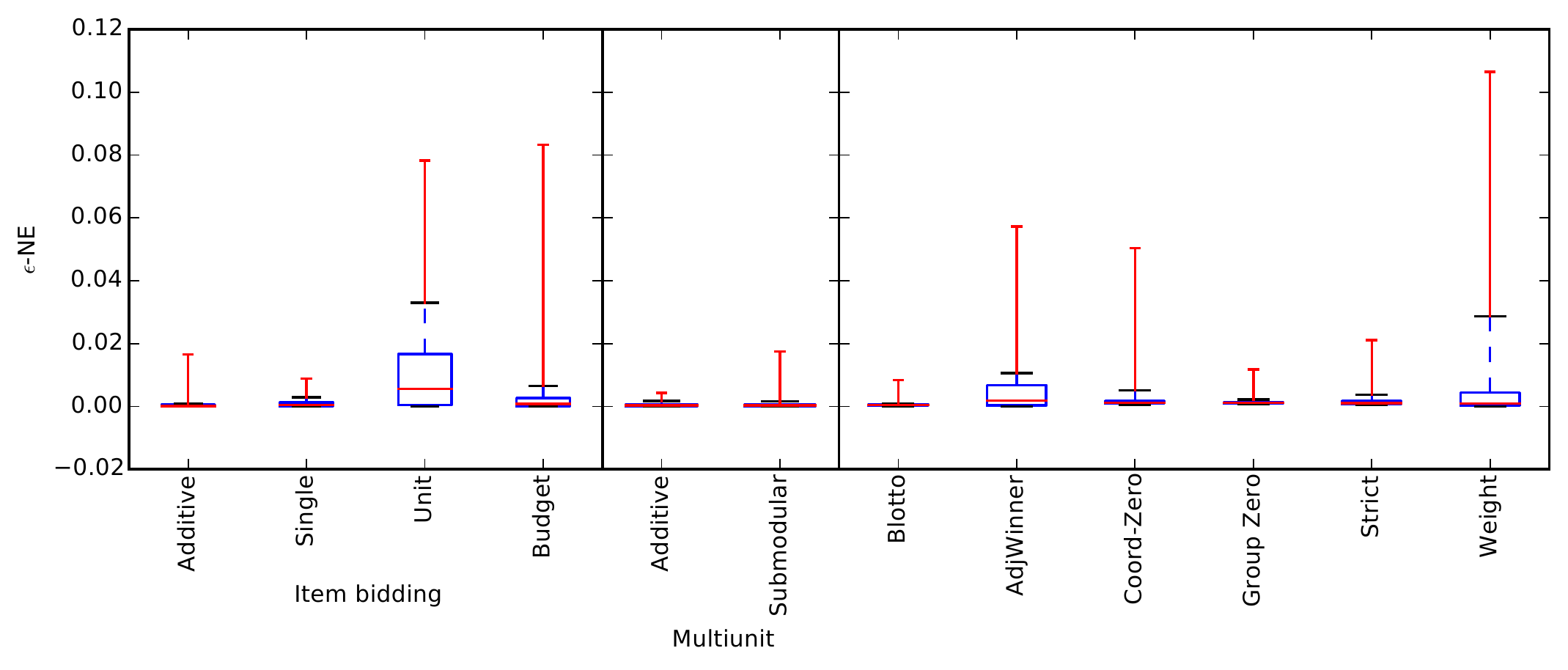}
\caption{Box and whisker plots showing the approximation quality of the
	approximate equilibria found by \descent with $\delta = 0.001$. The results
	show that \descent almost always finds a high quality approximate
	equilibrium. It can be seen that on many classes of games, even the worst
	approximation quality over all test cases is very good.} 
	\label{fig:box}
\end{figure*}

Figure~\ref{fig:box} shows a box and whisker plot for the quality of approximate
equilibrium found by accurate \descent variant.  It can be seen
that even the worst performance of the algorithm is relatively good for several
classes of game. The overall worst approximate equilibrium was a $0.1065$-NE that was found on a
weighted cooperation game. While this is far larger than the average
performance, it is still much better than best-known theoretical upper bound of~$0.5$.

We now make more detailed observations about the specific classes of games that
we tested. For auctions, one interesting observation is that on certain classes
of auctions \lemke will often find a pure Nash equilibrium. This is shown in the
``\% Pure column'' of Table~\ref{tbl:auction}. This phenomenon is particularly
prevalent for second-price auctions, where in some cases we found that \lemke
always finds a pure, and in these cases it does so in a very small amount of
time.

%\begin{figure*}
%\centering
%\begin{subfigure}{.4\textwidth}
%\includegraphics[page=1,width=\linewidth]{plots/item_ties.pdf}
%\end{subfigure}
%\begin{subfigure}{.4\textwidth}
%\includegraphics[page=2,width=\linewidth]{plots/item_ties.pdf}
%\end{subfigure}
%\caption{Plots showing the performance on \lemke on item bidding auctions on first price, second
%price and all-pay payment methods, on the three different tie resolution methods.}
%\end{figure*}

We also found that the tie-breaking rule used
in the auction can have a huge impact on the time that \lemke takes to find an
exact equilibrium. Table~\ref{tbl:ties} shows the performance of the algorithm
on otherwise identical auctions with different tie-breaking rules. It can be
seen that resolving ties deterministically makes the game much easier to solve
than resolving ties randomly.

Finally, we discuss the results in Table~\ref{tbl:ls} for  \descent
with and without line search. It
can be seen that, without line search, the \descent algorithm with $\delta=0.001$ 
is often slower than \lemke, and that using line search greatly
speeds it up (and so in our other results we always used line search). 
Interestingly, the line-search variant of
the algorithm also finds better quality approximate equilibria;
it would be interesting to understand why.

\section{Conclusions}

In this paper we extensively studied the performance of two algorithms
for computing a sample equilibria of polymatrix games.
Both algorithms produce
good results for most of the test instances, even though many were drawn from
theoretically hard classes. More specifically, we saw that
combinatorial auctions with two bidders are relatively easy to solve. This
raises the natural question whether we can derive efficient algorithms for
auctions with two, or a constant number of players. Furthermore, we saw that 
tie resolution significantly affects the difficulty of the auctions (see~\cite{CP14} 
for a discussion of this issue in a theoretical context).

In all of our experiments \descent produced \eps-NE far better than the
(best-known) 0.5 theoretical worst-case guarantee, which is not known to be tight.  
So it would be interesting to understand if
this good performance is due to the nature of the games we studied or if there is a better
theoretical analysis. 
In~\cite{FIS15}, a genetic algorithm was used to construct a bimatrix 
game for which 
the Tsaknakis and Spirakis (TS) algorithm computes an 0.3393-NE, which shows the 
analysis of the TS algorithm is essentially tight.  
Since bimatrix games are a
special case of polymatrix games, this gives a lower bound of 0.03393 for the 
best-possible approximation guarantee for \descent in polymatrix games. 
Can a better lower bound, closer to the 0.5 upper bound, be found? We believe that it
should be easier to construct a bad game for the \descent algorithm compared to
the TS algorithm,  because \descent computes a single strategy profile,
whereas TS computes three profiles (one by descent, and then two further
profiles are derived from that one) and then chooses the best one.

\smallskip

\noindent \textbf{Acknowledgements.} The last author is supported by EPSRC grant EP/L011018/1.

% For AAMAS-2016, as references are unlimited but appendices must fit within
% 8 pages, the References section must come after the appendices (if any)
\clearpage
%\bibliographystyle{abbrv}
%\bibliography{references}  % sigproc.bib is the name of the Bibliography in this case

\newpage
\appendix
\section{Line Search}
\label{app:linesearch}

%This procedure is
%done by representing the $\epsilon$-NE of the new point in the \descent method
%$x + \alpha(x' - x)$ as a piecewise function in terms of $\alpha$, which is then
%evaluated for different values of $\alpha \in (0, 1]$, with the aim of finding
%an $\alpha$ which minimizes the $\epsilon$-NE. Prior to performing the search,
%$\alpha = \frac{\delta}{\delta + 2}$ which has been proven to get a new point
%with a better approximation value than $x$, the current point. 

As noted in the main text, the line search procedure for \descent greatly
affects the running time of the running time of the algorithm. In this section,
we give some results that describe this impact.

Recall that the line search procedure selects a number of equally spaced points
for $\alpha$ between $0$ and $1$, and the key decision is how many points to
check. In order to study the effects of the number of points on line search has
on the algorithm, we took the average over 100 random instances on
Coordination/Zero-sum games on cycles and grids as well as Weighted Zero-sum
games on Cycles.

\begin{figure*}[th]
\centering
\begin{tabular}{cc}
\includegraphics[page=1,width=.4\linewidth]{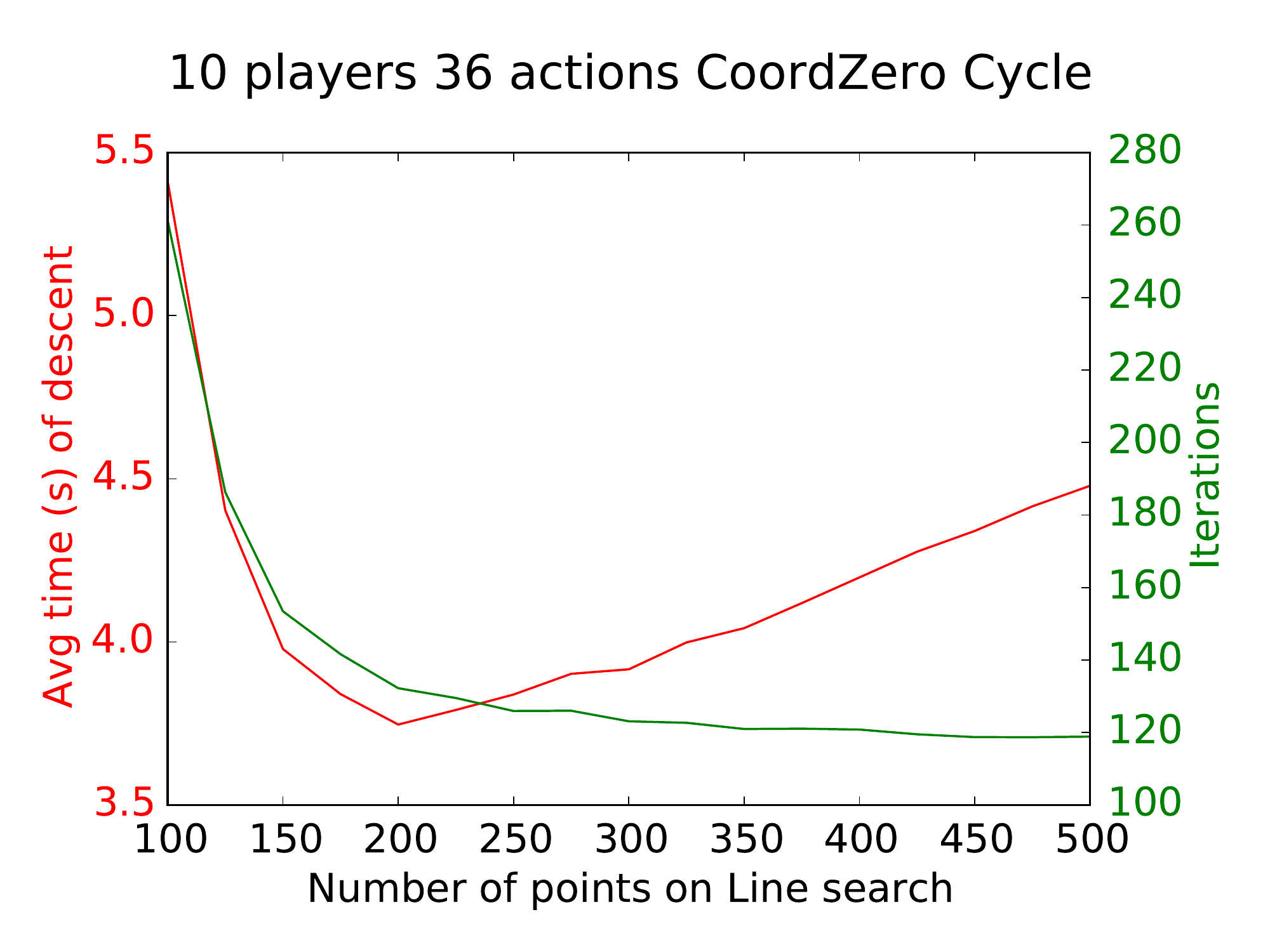} &

\includegraphics[page=3,width=.4\linewidth]{plots/descent_iter.pdf}\\

\includegraphics[page=5,width=.4\linewidth]{plots/descent_iter.pdf} &

\includegraphics[page=7,width=.4\linewidth]{plots/descent_iter.pdf}\\
\end{tabular}
\caption{Plots showing the effect of varying the number of points used in line search on the number of
iterations and computation time of \descent.}
\label{fig:linesearch}
\end{figure*}

Figure~\ref{fig:linesearch} shows the results. The red lines show how the
overall running time of the algorithm varies depending on the number of point
checked in each iteration, while the green lines show how the number of
iterations change. As one might expect, the number of iterations tends to
decrease as we increase the number of points checked in each iteration. In other
words, checking more points increases the amount of progress that each iteration
makes. However, checking more points in each iteration also adds more
computational cost, and eventually the overall running time begins to rise,
which indicates that the cost of checking more points is too high relative to
the extra progress that is made.

%, the number of points on the line
%search affects the performance of the algorithm. Larger search spaces tend to
%result in a reduction in the number of iterations, however there is a trade-off
%as this gets more expensive to perform the search and this ends up slowing down
%the algorithm. This point tends to differ based on the size of the game, with
%more points needed for larger games.

\section{Game size: Players vs Actions}
\label{app:playersvactions}

The size of a polymatrix game increases with more players and with more actions.
In general, we found that the number of players had a much greater effect on the
running time of the algorithms that the number of actions. Table~\ref{tab:playersactiontradeoff} 
illustrates this point.

\begin{table*}[htp]{
\centering
\begin{tabular}{|c|c|c|c|c||c|c||c|c||c|c|c|}
\multicolumn{5}{|c||}{Games} & \multicolumn{2}{c||}{\lemke} & \multicolumn{2}{c||}{\descent 0.1 LS} &
\multicolumn{3}{c|}{\descent 0.001 LS} \\\hline
Players & Actions& \# Payoff & LCP & $p$ & Time & \% T & Time & \eps & Time & \eps & \% T\\
\hhline{|=|=|=|=|=||=|=||=|=||=|=|=|}
 & & 992250 & 1060 & 2 & 3.113 & 0.0 & 0.301 & 7.152e-04 & 12.586 & 8.502e-04 & 0.0 \\\cline{3-12}
10 & 105 & 992250 & 1060 & 3 & 1.933 & 0.0 & 0.262 & 5.853e-04 & 13.704 & 8.381e-04 & 0.0 \\\cline{3-12}
 & & 992250 & 1060 & 5 & 1.430 & 0.0 & 0.198 & 6.931e-04 & 15.256 & 7.425e-04 & 0.0 \\\cline{3-12}
\hhline{|=|=|=|=|=||=|=||=|=||=|=|=|}
 & & 995000 & 1200 & 2 & 194.233 & 8.0 & 8.455 & 1.446e-02 & 86.116 & 9.603e-04 & 0.0 \\\cline{3-12}
200 & 5 & 995000 & 1200 & 3 & 410.118 & 38.0 & 6.551 & 1.909e-02 & 73.485 & 1.047e-03 & 0.0 \\\cline{3-12}
 & & 995000 & 1200 & 5 & 552.583 & 81.0 & 4.957 & 2.585e-02 & 64.676 & 1.130e-03 & 0.0 \\\hline\end{tabular}

\caption{Table showing Weighted Cooperation games on Complete graphs relatively 
close in size, comparing the case of a large number of players and few actions 
with the case of a few players with larger action sets. $\% T$ is the proportion of the timed 
out instances.}
\label{tab:playersactiontradeoff}
}
\end{table*}
%
% ACM needs 'a single self-contained file'!
%\balancecolumns % GM June 2007
% That's all folks!
\end{document}